\documentclass[a4paper]{article}
\usepackage[dvips]{graphicx}
\usepackage{amssymb}
\usepackage{latexsym}
\usepackage{bm}

\newcommand{\I}{\mathrm{i}}

\newcommand{\qbar}{\ensuremath{\overline{q}}}
\newcommand{\be}{\begin{equation}}
\newcommand{\ee}{\end{equation}}

\begin{document}

\date{\today}
\title{Chiral symmetry breaking and the spin content of the
$\rho$ and $\rho'$ mesons}
\author{\bf L.~Ya.~Glozman\footnote{leonid.glozman@uni-graz.at},
C.~B.~Lang\footnote{christian.lang@uni-graz.at}, 
and M.~Limmer\footnote{markus.limmer@uni-graz.at}}
\maketitle
\begin{center}
Institut f\"ur Physik, FB Theoretische Physik\\
Universit\"at Graz, A-8010 Graz, Austria\\
~\\
\end{center}
\begin{abstract}
Using interpolators with different $SU(2)_L \times SU(2)_R$ transformation
properties we study the chiral symmetry and spin contents of the $\rho$- and
$\rho'$-mesons in lattice simulations with dynamical quarks. A ratio of couplings of the
$\qbar\gamma^i{\tau}q$ and $\qbar\sigma^{0i}{\tau}q$ interpolators to a
given meson state at different resolution scales tells one about the degree of chiral symmetry
breaking in the meson wave function at these scales. Using a Gaussian  gauge
invariant smearing of the quark fields in the interpolators, we are able to
extract the chiral content of mesons up to the infrared resolution of $\sim1$
fm. In the ground state $\rho$ meson the chiral symmetry is strongly broken with
comparable contributions of both the $(0,\,1) + (1,\,0)$ and  $(1/2,\, 1/2)_b$
chiral representations with the former being the leading contribution. In
contrast, in the $\rho'$ meson the degree of chiral symmetry breaking is
manifestly smaller and the leading representation is $(1/2,\, 1/2)_b$. Using a
unitary transformation from the chiral basis to the $^{2S +1}L_J$ basis, we are
able to define and measure the angular momentum content of mesons in the rest
frame. This definition is different from the traditional one which uses parton
distributions in the infinite momentum frame. The $\rho$ meson is practically a 
$^3S_1$ state with no obvious trace of a ``spin crisis". The $\rho'$ meson has a
sizeable contribution of the $^3D_1$ wave, which implies that the $\rho'$ meson
cannot be considered as a pure radial excitation of the $\rho$ meson.
\end{abstract}

\section{Introduction}

The structure of hadrons in the infrared is a challenging topic. At low resolution
scales (i..e., large distances $\mathcal{O}(1\;\textrm{fm})$) both, confinement and chiral symmetry breaking, are crucial phenomena. They
influence mass and angular momentum generation of hadrons. These phenomena are 
of primary interest both theoretically and
experimentally. To understand physics at these deeply nonperturbative scales one
needs direct information about the chiral and the angular momentum content of hadrons.

Such information can be obtained from dynamical lattice simulations. Using a set
of interpolators that form a complete basis with respect to the $SU(2)_R \times
SU(2)_L$ transformations, one is able to define in a gauge invariant manner and
measure the chiral content of mesons at different resolution scales \cite{GLL1}.
One basically measures a ratio of couplings of different interpolators to a
given hadron. Such a ratio tells us something about chiral
symmetry breaking in a hadron wave function. If chiral symmetry were unbroken in
a hadron, then only interpolators with definite chiral transformation properties
would couple to this hadron. Chiral symmetry breaking in a hadron would imply
that interpolators with different chiral transformation properties would create
this hadron from the vacuum.

The output of the two-flavor dynamical simulations with $m_{AWI} \sim 15 - 30$
MeV \cite{GLL1} was that at the resolution scales  $0.15 - 0.6$ fm the $\rho$ meson
is approximately a 55\% - 45\% mixture of the two possible chiral
representations $(0,1) + (1,0)$ and $(1/2,1/2)_b$. Given a unitary
transformation from the quark-antiquark chiral basis to the $^{2S+1}L_J$ basis
in the rest frame \cite{GN}, we were able to extract the angular momentum content
of the $\rho$ meson in the rest frame \cite{GLL1}. The result was that the
$\rho$ meson at the scales $0.15 - 0.6$ fm is approximately a $^3S_1$ state with a
tiny contribution of a $^3D_1$ wave. In this definition of
total angular momentum (in the rest frame) there is no ``spin crisis'', at least
for the $\rho$ meson.
This definition of the spin content of a hadron is very different from the
traditional one. The latter relies on the parton distributions in the infinite
momentum frame extracted from the deep inelastic scattering with polarization
\cite{EMC1}. Accordingly only about 30\% spin of the nucleon is carried by the
spins of valence quarks \cite{EMC2}, which gave rise to the term ``spin crisis".
Similar results within this same definition
are obtained on the lattice both for the nucleon  and mesons, for a review and
references see \cite{Hag}. Then a natural question is which of these two
definitions does reflect the spin content of a hadron?

In  \cite{GLL2} we have also measured the chiral and angular momentum content
of the first excitation of the $\rho$ meson, $\rho' \equiv \rho(1450)$. As compared
to the ground state $\rho$, we have observed a very different dependence of the
chiral and angular momentum content on the resolution scale. In particular, we
have found weaker chiral symmetry breaking in the $\rho'$ state with the
$(1/2,1/2)_b$ representation being the leading  one in the infrared. We have also
found a significant contribution of the $^3D_1$ wave in the $\rho'$ wave
function. This means that the $\rho'$ cannot be considered as a pure radial
excitation of  the $\rho$ meson.

In these studies only two different resolution scales were used. This did not
allow to reliably extrapolate the results up to the resolution scale of the
excited hadron size, $\sim 1$ fm. In the present paper we extend our correlation
matrix from $4 \times 4$ to $6 \times 6$ by providing three different Gaussian
smearings of the quark fields in the vector and tensor interpolators. This allows us
to discuss the chiral symmetry and angular momentum
content of both $\rho$ and $\rho'$ mesons in the infrared region of 1 fm, where
mass is generated.
 
\section{Theoretical foundations of the method}

In this paper we study the chiral and angular contents of the $\rho$-mesons,
consequently we restrict our discussions specifically to the $I=1,
J^{PC}=1^{--}$ states. But the formalism is generic and can be used for mesons
with other quantum numbers as well.

A chiral classification of some  interpolators is performed in Ref.\ \cite{Cohen:1996sb}
and a full classification  of the quark-antiquark states as well as of the
corresponding interpolators is done in Ref.\ \cite{G1}. In the case of the $I=1,
J^{PC}=1^{--}$ states there are two allowed chiral representations, $(0,1) +
(1,0)$ and $(1/2,1/2)_b$. The  state that transforms as $(0,1) + (1,0)$ can be
created from the vacuum by the vector current, 
\begin{equation}
O_\rho^V(x)  = \qbar(x)\, \gamma^i \vec \tau\, q(x)\;,
\label{rV}
\end{equation}
and the state that belongs to the $(1/2,1/2)_b$ representation 
can be created by the pseudotensor operator,
\begin{equation}
O_\rho^T(x)  = \qbar(x)\, \sigma^{0i} \vec \tau\, q(x)\;.
\label{rT}
\end{equation}
The chiral partner of the first operator is the axial vector current,
\begin{equation}
O_{a_1}(x)  = \qbar(x)\, \gamma^i \gamma^5 \vec \tau\, q(x)\;,
\label{a1}
\end{equation}
that creates from the vacuum the $a_1$ states, $I=1, J^{PC}=1^{++}$. The
chiral partner of the second operator is the  operator
\begin{equation}
O_{h_1}(x)  = \varepsilon ^{ijk} \qbar(x)\, \sigma^{jk} \, q(x)\;,
\label{h1}
\end{equation}
that couples to the $I=0, J^{PC}=1^{+-}$ $h_1$ mesons.

It is well established in quenched \cite{LQ} and dynamical 
\cite{GLL1,GLL2,LD} lattice simulations that the ground and excited states
of the $\rho$ meson can be created from the vacuum by both the vector and
pseudotensor operators. This fact by itself means that chiral symmetry is broken
in the vacuum and in the physical states, and these states are mixtures of
these two representations \cite{GLL1,GLL2}. 

The chiral basis in the quark-antiquark system is a complete one and can be
connected to the complete angular momentum basis in the rest frame via the
unitary transformation \cite{GN}
\begin{equation}\label{unitary_1}
\left(
\begin{array}{l}
|(0,1)\oplus(1,0);1 ~ 1^{--}\rangle\cr
|(1/2,1/2)_b;1 ~ 1^{--}\rangle
\end{array}
\right) = U\cdot
\left(
\begin{array}{l}
|1;{}^3S_1\rangle\cr
|1;{}^3D_1\rangle
\end{array}
\right)
\end{equation}
with 
\begin{equation}\label{unitary_2}
U=
\left(
\begin{array}{cc}
\sqrt{\frac23} & \sqrt{\frac13} \cr
\sqrt{\frac13} & -\sqrt{\frac23} 
\end{array}
\right)\;.
\end{equation}
Consequently, if we know the mixture of the two allowed chiral representations in
a physical state, we are also able to obtain the angular momentum content of this
state in the rest frame.

In particular, we can answer a question whether or not a spin of a meson
is carried by spins of its valence quarks in the rest frame. This definition
of the spin content is different from the traditional one that relies on
the parton distributions in the infinite momentum frame. According to the
latter only about 30\% of the nucleon spin is carried by the valence quarks,
which has been referred to as ``spin crisis". Hence we can compare a spin content
of a meson obtained according to our definition with the spin content
extracted from the parton distributions in the infinite momentum frame.

\section{Reconstruction of the coupling constants with the variational method}

In order to resolve a few subsequent physical states with the same quantum
numbers one has to choose a convenient set of operators $O_i$ with the same
quantum numbers and a significant overlap with the physical states,  and
calculate a cross-correlation matrix at zero spatial momentum  (i.e., in the
rest frame) \cite{VAR},
\begin{equation}\label{corr_inf}
C(t)_{ij}=\langle O_i(t)O_j^\dagger(0)\rangle=\sum_{n=1}^\infty a_i^{(n)} a_j^{(n)*} 
\mathrm{e}^{-E^{(n)} t}\;,
\end{equation}
with the coefficients giving the overlap of the  operators with the
physical state,
\begin{equation}\label{eq_w_f}
a_i^{(n)}=\langle 0| O_i|n\rangle\;.
\end{equation}

With a set of operators spanning a complete and orthogonal basis with  respect
to some symmetry group, these overlaps (coupling constants) give the complete
information about symmetry breaking. The interpolating composite operators $O_i$
are not normalized on the lattice
and consequently the absolute values of the coupling constants $a_i^{(n)}$
cannot be obtained. However, a ratio of the couplings is a well defined quantity
and can be computed as \cite{GLL1}
 \begin{equation}\label{ratio_op_comp}
\frac{a_i^{(n)}}{a_k^{(n)}}=
\frac{\widehat C(t)_{ij} u_j^{(n)}}{\widehat C(t)_{kj} u_j^{(n)}}\;.
\end{equation}
Here $\widehat C$ is the cross-correlation matrix from (\ref{corr_inf}), 
a  sum is implied for the index $j$ on the right-hand side and $u_j^{(n)}$ are the eigenvectors obtained from the generalized eigenvalue problem,
\begin{equation}\label{gev_1}
\widehat C(t)_{ij} u_j^{(n)} =\lambda^{(n)}(t,t_0)\widehat 
C(t_0)_{ij} u_j^{(n)}\;,
\end{equation}
with $t_0$ being some normalization point in Euclidean time. 

In our calculation a set of interpolators $O_i$ complete with respect to
$SU(2)_L \times SU(2)_R$ and the angular momentum basis consists of the vector
(\ref{rV}) and pseudotensor (\ref{rT}) operators. However, there is an infinite
amount of nonlocal  operators with the same chiral and angular momentum
structure like (\ref{rV}) and (\ref{rT}) but with different radial spatial form.
We want to construct these nonlocal operators in such a way that each of them
would probe the hadron structure at a given physical resolution scale.

In the continuum the corresponding amplitudes are given as
\begin{eqnarray}
 \langle 0 | \qbar(0) \gamma^\mu q(0) | V(p; \lambda)\rangle &=& 
 m_\rho f_\rho^V e^\mu_\lambda\;,
\label{rhoV}\\
 \langle 0| \left(\qbar(0) \sigma^{\alpha \beta} q(0)\right)(\mu) | V(p; \lambda)\rangle &= &
 \I f_\rho^T(\mu) e^\mu_\lambda 
 (e^\alpha_\lambda p^\beta -  e^\beta_\lambda p^\alpha)\;,
\label{rhoT}
\end{eqnarray}
where   $V(p; \lambda)$ is the vector meson state with the mass $m_\rho$,
momentum $p$ and polarization $\lambda$. The vector current is conserved,
consequently the vector coupling constant $f_\rho^V$ is scale-independent. The
pseudotensor ``current'' is not conserved and is subject to a nonzero anomalous
dimension. Consequently the pseudotensor coupling $ f_\rho^T(\mu)$ manifestly
depends on the  scale $\mu$. In the rest frame the ratio
\begin{equation} 
\frac{f_\rho^V}{f_\rho^T(\mu)} =  \frac
 {\langle 0 | \qbar(0) \gamma^i q(0) | V(\lambda)\rangle}
 {\langle 0 |\left(\qbar(0) \sigma^{0i} q(0)\right)(\mu) | V(\lambda)\rangle}
 \label{rhoV/rhoT}
\end{equation}
coincides with the ratio of matrix elements (\ref{ratio_op_comp}) with $i \equiv
V; ~ k \equiv T$.

\section{Physical resolution scale}

We want to probe the hadron structure at infrared scales, where mass is
generated. The hadron interpolators that create and annihilate the hadrons are built from
quark fields.
With the local lattice interpolators of type $\qbar(x) \Gamma q(x)$ we study the hadron structure at the
scale given by the lattice spacing $a$. Given a reasonably small value of $a$ we
can fix a smaller resolution (larger size) of our probe 
by a gauge invariant smearing of the quark
fields in the interpolator. Namely, we smear the quark fields in the source and
sink in spatial directions  with a Gaussian profile of the size $R$ . Technically this Gaussian
type of smearing is achieved by  Jacobi  smearing \cite{Jac}. For examples of the resulting
quark profiles see, e.g., \cite{Burch:2006dg}.

The eigenvectors of the cross correlation matrix then give us information on the
contribution of the various interpolators (with different Dirac structure and built
from differently smeared quark sources) to the physical state.

Then, even in the continuum limit $a \rightarrow 0$ we may probe the hadron structure
at a scale fixed by $R$. Such a definition of the resolution is similar to the
experimental one, where an external probe is sensitive only to the quark fields
(it is blind to gluonic fields) at a resolution that is  determined by the
typical momentum transfer in spatial directions.

\section{Lattice details and choices of correlation matrix}

\begin{table}
\begin{center}
\begin{tabular}{ccccccccc}
\hline
\hline
Set & $\beta_{LW}$ & $a\,m_0$ & \#{conf} & $a$ [fm] & $m_\pi$ [MeV] & $m_\rho$ [MeV] & $m_{\rho'}$ [MeV]\\
\hline
A\phantom{1} & 4.70 & -0.050 & 200 & 0.1507(17) & 526(7) & 911(11) & 1964(182)\\
B1           & 4.65 & -0.060 & 300 & 0.1500(12) & 469(5) & 870(10) & 1676(106)\\
B2           & 4.65 & -0.070 & 200 & 0.1406(11) & 296(6) & 819(18) & 1600(181)\\
C\phantom{1} & 4.58 & -0.077 & 300 & 0.1440(12) & 323(5) & 795(15) & 1580(159)\\
\hline
\hline
\end{tabular}
\caption{\label{tab:sim}
Specification of the data used here; for the gauge coupling only the
leading value $\beta_{LW}$ is given, $m_0$ denotes the bare mass parameter of
the CI action. Further details on the action, the simulation and the
determination of the lattice spacing and the $\pi$- and $\rho$-masses are found
in \cite{Gattringer:2008vj,Engel:2010my}.}
\end{center}
\end{table}

\begin{table}
\begin{center}
\begin{tabular}{cccc}
\hline
\hline
Set & $R_n$ [fm] & $R_w$ [fm] & $R_{uw}$ [fm]\\
\hline
A\phantom{1}  & 0.36 & 0.67 & --\\
B1            & 0.34 & 0.69 & 0.81\\
B2            & 0.34 & 0.66 & 0.85\\
C\phantom{1}  & 0.33 & 0.66 & --\\
\hline
\hline
\end{tabular}
\caption{\label{tab:radii}
Specification of the smearing radii $R$.}
\end{center}
\end{table}

\begin{figure}[t]
\begin{center}
\includegraphics[height=8cm,clip]{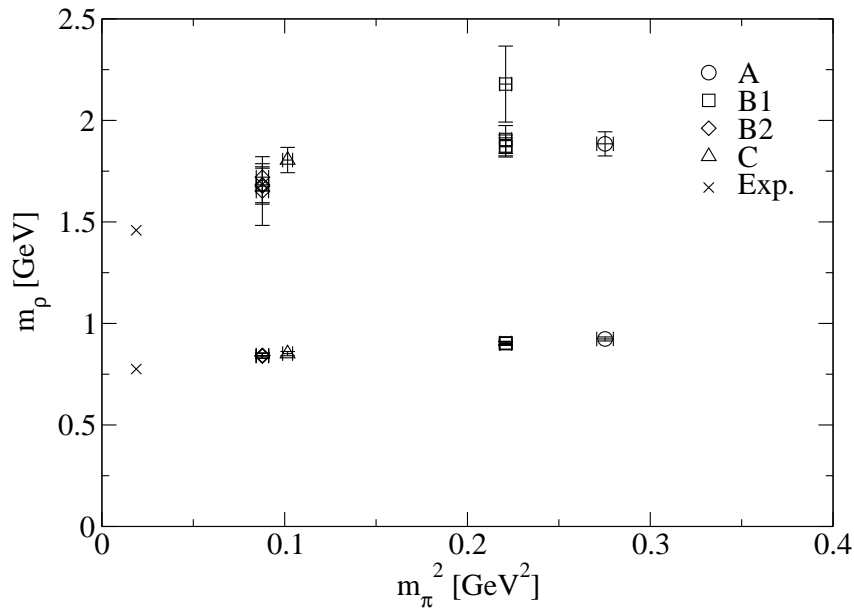}
\end{center}
\caption{\label{fig:masses}
The masses of both $\rho$ and $\rho'$ states extracted from
different $4 \times 4$ and $6 \times 6$ correlation matrices. The crosses indicate the mass
values from experiments.} 
\end{figure}

In our study we use Chirally Improved fermions \cite{CI} and the
L\"uscher-Weisz gauge action \cite{LW}. The lattice size is $16^3 \times 32$. We
use dynamical gauge configurations with two mass-degenerate light quarks.
With the lattice spacing $\approx 0.15$ fm the spatial volume of the  lattice is
$\approx 2.4^3$ fm$^3$. For the ground states and some their first excitations
such a volume turns out to be sufficient to get  approximately correct masses of
hadrons  in the physical limit \cite{Engel:2010my}, though it is certainly too small to
consider higher excitations. In our present study we limit ourselves to the
$\rho$ and $\rho'$ states. For  details on the simulation we refer the reader to the
Table \ref{tab:sim} and to \cite{Gattringer:2008vj,Engel:2010my} .

We use three different smearing radii $R$ for the quark fields in the source and
sink, see Table \ref{tab:radii}. The ``narrow" smearing width (index $n$)  varies between
0.33 and 0.36 fm, depending on the set of configurations. The ``wide" smearing
radius (index $w$)  lies between 0.66 and 0.69 fm and the ``ultrawide" one is 0.81 -- 
0.85 fm (index $uw$).  Hence we can study the hadron structure at  resolutions 
0.33 fm  -- 0.85 fm and will be able to extrapolate the results up a resolution of 
$\mathcal{O}(1\;\textrm{fm})$.

We have the following set of operators: 
\begin{eqnarray}
O^V_n=\overline u_n \gamma^i d_n\;,\;\;
&O^V_w=\overline u_w \gamma^i d_w\;,\;\;
&O^V_{uw}=\overline u_{uw} \gamma^i  d_{uw}\;,\;\;\nonumber\\
O^T_n=\overline u_n \gamma^t \gamma^i  d_n\;,\;\;
&O^T_w=\overline u_w \gamma^t \gamma^i  d_w\;,
&O^T_{uw}=\overline u_{uw} \gamma^t \gamma^i  d_{uw}\;,
\end{eqnarray}
where $\gamma^i$ is one of the spatial Dirac matrices and $\gamma_t$ is the
$\gamma$-matrix in (Euclidean) time direction. For the sets A and C we
construct $4 \times 4$ correlation matrices (i.e., with both vector and
pseudotensor interpolators using narrow and wide smearing radii), while for
the sets B1 and B2 we study the $6 \times 6$ correlation matrix (with
narrow, wide and ultrawide smearings for both vector and pseudotensor
operators) as well as different possible $4 \times 4$ sub-matrices.

For the parameters of the simulation the $\rho$ mass is below the p wave decay 
energy. As has been discussed in \cite{Engel:2010my} no coupling to the $\pi\pi$
channel is observed, which may be due to the fact that no meson-meson interpolator
has been explicitly included in the set. We thus may identify the second lowest 
observed energy level with the $\rho'$ (see also \cite{Pr}).
The masses of both $\rho$ and $\rho'$ states, extracted from different sets and
correlation matrices, are shown in Fig.\ \ref{fig:masses}.

\section{Chiral symmetry breaking and the angular momentum content
of $\rho$ and $\rho'$ mesons}

\begin{figure}[t]
\begin{center}
\includegraphics*[height=8cm,clip]{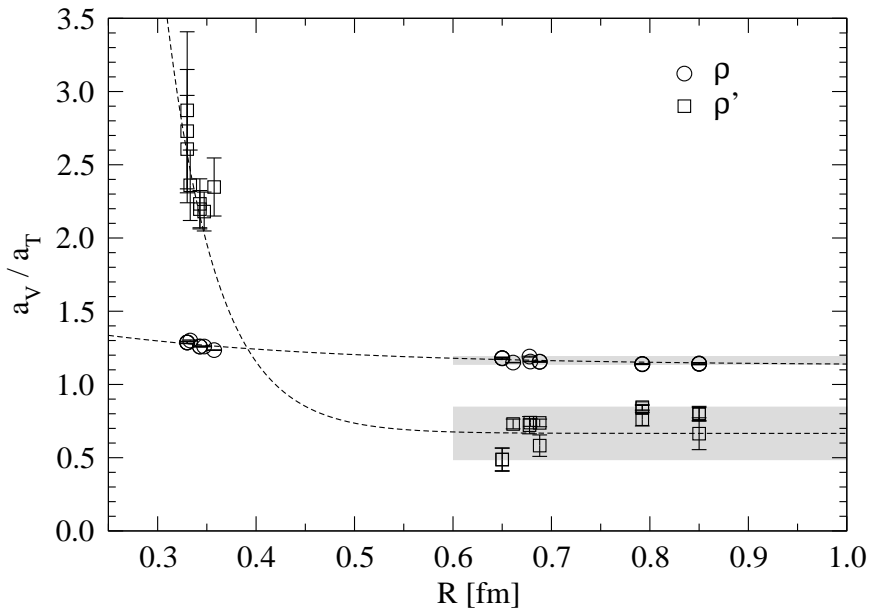}
\end{center}
\caption{\label{fig:all}
A ratio of the vector to the pseudotensor couplings versus a resolution scale $R$, as
extracted from all  $4 \times 4$ and $6 \times 6$ correlation matrices.
Broken lines are drawn only to guide the eye.}
\end{figure}

A measure of  chiral symmetry breaking in $\rho$ and $\rho'$ states at some
resolution $R$ is given by the ratio $a_V/a_T$, which is the same as the ratio
(\ref{rhoV/rhoT}), obtained at a given resolution scale. At $\mu \to
\infty$ (i.e., at $a \to 0$, $R\to 0$) this ratio is divergent, because the
pseudotensor operator decouples from the physical state in the asymptotic
freedom regime. This can be understood in two ways. In the asymptotic freedom
regime chiral symmetry is not broken, hence only one of the two interpolators
(which have different chiral transformation properties) can couple to the state.
The vector current is conserved and the constants $f^V$ are scale-independent.
The pseudotensor ``current'' is not conserved and in the asymptotic freedom regime
its coupling approaches zero \cite{LQ}. Indeed, in our previous study \cite{GLL1} 
we did observe
this behavior  for the $\rho$ meson towards the ultraviolet regime. However, the
way how this ratio approaches the ultraviolet regime for different physical
states is a priori unknown and explicitly depends on the hadron wave functions.

The results for this ratio for both the $\rho$ and $\rho'$ states
obtained from different $4 \times 4$ and $6 \times 6$ correlation matrices
are consistent with each other. Fig. \ref{fig:all}
shows the results for  $\rho$ and $\rho'$ from both sets of correlation matrices.
A complete set of operators would
include all possible smearing radii $R$ and the correlation matrix would be of
infinite dimension. Given the consistency of the results
fror  $4 \times 4$ and $6 \times 6$ we can deduce
reliable physical information already from the present  results.

In particular, we clearly see that the ratio for the $\rho$ and $\rho'$ mesons
approaches the ultraviolet regime in a very different manner. For the $\rho'$
state the pseudotensor operator decouples much faster towards the ultraviolet
than for the $\rho$ meson. This demonstrates that the wave functions of these
two states are significantly different. Since the ratio reflects a degree of
chiral symmetry breaking at different scales, this symmetry breaking is very
different for both states.

The ratio $a_V/a_T$, which is a ratio of two possible chiral representations
in a hadron wave function, also defines an
angular momentum decomposition of a state via the unitary transformation
(\ref{unitary_1}), (\ref{unitary_2}). In particular, in the asymptotic freedom
regime, where only the $(0,1)+(1,0)$ representation couples, the angular
momentum content of $\rho$ mesons is fixed to be $\sqrt{2/3}\,|^3S_1\rangle +
\sqrt{1/3}\,|^3D_1\rangle$. Deviation from this superposition of the $S$- and
$D$-waves in a state towards the  infrared regime is due to chiral symmetry
breaking in this state.  From Fig.\ \ref{fig:all} we clearly see  that the ratio is very
different for both states at all possible scales. Hence chiral symmetry 
breaking as well as the angular momentum generation are also very different for
both states. This difference shows that the $\rho'$ state cannot be considered
as simply a radial excitation of the $\rho$ meson, since the latter would
require that their angular momentum content is the same at different
resolutions.

Existing data at the resolutions $R \sim 0.65 - 0.85$ fm allows us to extrapolate
results up to a resolution of 1 fm. A tentative extrapolation with
uncertainties is represented by a shadowed area on Fig.\ \ref{fig:all}. Given that the physical
size of the $\rho$ state is of the order $0.7 - 0.8$ fm, as could be
deduced from the experimental charge radii of the nucleon and $\rho$-meson,
and for the excited $\rho'$
meson it should be of the order of 1 fm, we can deduce  chiral symmetry breaking
and the angular momentum content of both states at the infrared scales of their
size.

For the ground state $\rho$ meson this ratio is  within $a^\rho_V/a^\rho_T = 1.14-1.19$
while for its first excitation  this ratio $a^{\rho'}_V/a^{\rho'}_T =
0.48-0.84$. Chiral symmetry is almost maximally broken (i.e., close to 1)
in the ground
state $\rho$, while  a degree of chiral symmetry breaking in the $\rho'$ state
is essentially smaller, which is consistent with  effective chiral restoration
in highly excited hadrons \cite{G1}. In the ground state $\rho$ the slightly
leading representation is  $(0,1)+(1,0)$, while in the excited $\rho'$ state a
leading chiral representation is $(1/2,1/2)_b$.

Given these ratios, we can obtain the angular momentum contents of both mesons
from (\ref{unitary_1}), (\ref{unitary_2}). For
the $\rho$ meson it is approximately  $0.99\,|^3S_1\rangle -
0.1\,|^3D_1\rangle$. Hence the ground state in the infrared is practically a
pure $^3S_1$ state with a tiny admixture of the $^3D_1$ wave. 

In contrast, 
in the excited $\rho$ meson there
is a sizeable contribution of the $^3D_1$ wave. In the
latter case the angular momentum content is between the following two lower and
upper bound values. For the lower bound it is $0.88\,|^3S_1\rangle -
0.48\,|^3D_1\rangle$ and for the  upper bound it is $0.97\,|^3S_1\rangle -
0.25\,|^3D_1\rangle$. This once again demonstrates that the first excitation of
the $\rho$ meson cannot be considered as a pure radial excitation of the ground state $\rho$. Obviously,
both radial and orbital degrees of freedom are excited which reflects yet
unknown dynamics of confinement and chiral symmetry breaking.

The fact that the angular momentum content of the $\rho$ meson is given by
the $^3S_1$ state means that according to a definition
used in our study there is no ``spin crisis" . The spin of the $\rho$ meson in the rest frame is carried by spins of its valence quarks dressed by gluons.
The gluonic field is important for the angular momentum generation, because it is this field that
provides chiral symmetry breaking and is responsible for most of the hadron mass. 
However, it is not clear to us whether it is possible 
to separate contributions of quarks and gluons in the highly non perturbative, confining
regime.

\section{Conclusions}

We summarize the most important implications of our study.
It is possible to define and measure a degree of chiral symmetry
breaking in mesons at different resolution scales.
We are able to define and measure in a gauge invariant manner the angular
momentum content of mesons in the rest frame.
The angular momentum content of hadrons is deeply connected with
chiral symmetry breaking in hadrons.
Chiral symmetry is strongly broken in the $\rho$ meson and its wave function
(Bethe-Salpeter amplitude) is approximately a 55\% - 45\% mixture of the chiral representations $(0,1)+(1,0)$
and $(1/2,1/2)_b$.
The angular momentum content of the $\rho$ meson is almost completely
represented by the $^3S_1$ partial wave at resolution scales of $0.15 - 1$ fm.
According to our definition of the spin content, the spin of the $\rho$
meson is carried by its valence quarks.
Chiral symmetry breaking in the excited $\rho'$ meson is weaker and the
leading chiral representation in this case is $(1/2,1/2)_b$.
There is a significant contribution of the $^3D_1$ wave in the $\rho'$
wave function and consequently the $\rho'$ meson cannot be considered as a
radial excitation of the $\rho$.

\paragraph{Acknowledgments.}

We gratefully acknowledge support of the grants P21970-N16 and DK W1203-N08 
of the Austrian
Science Fund FWF and the DFG project SFB/TR-55. The calculations have been performed
on the SGI Altix 4700 of the Leibniz-Rechenzentrum Munich and on the local
clusters at the University of Graz.

\end{document}